# Shape and effective spring constant of liquid interfaces probed at the nanometer scale: finite size effects

Julien Dupré de Baubigny,[1,2] Michael Benzaquen,[3] Laure Fabié,[1,2] Mathieu Delmas,[1] Jean-Pierre Aimé,[4] Marc Legros,[1] Thierry Ondarçuhu[1*]

[1] CEMES-CNRS, UPR 8011, 29 rue Jeanne Marvig, 31055 TOULOUSE

[2] Université de Toulouse, 29 rue Jeanne Marvig, 31055 TOULOUSE

[3] UMR CNRS 7083 Gulliver, ESPCI ParisTech, PSL Research University, 10 rue Vauquelin, 75005 PARIS

[4] CBMN, CNRS UMR 5248, 2 rue Escarpit, 33600 PESSAC

## ABSTRACT

We investigate the shape and mechanical properties of liquid interfaces down to nanometer scale by atomic force microscopy (AFM) and scanning electron microscopy (SEM) combined with in situ micromanipulation techniques. In both cases, the interface is probed with a 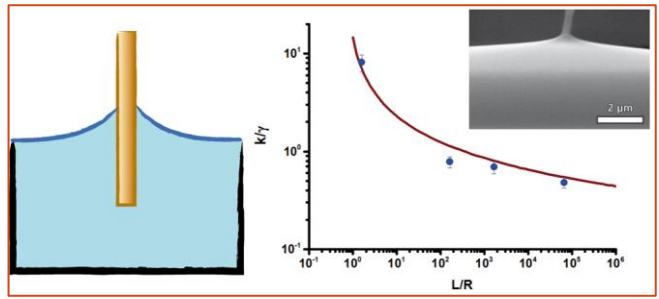 cylindrical nanofiber with radius $R$ of the order of 25-100 nm. The effective spring constant of the nanomeniscus oscillated around its equilibrium position is determined by static and frequency-modulation (FM) AFM modes. In the case of an unbounded meniscus, we find that the effective spring constant $k$ is proportional to the surface tension $\gamma$ of the liquid through $k = (0.51\pm0.06)\ \gamma$, regardless of the excitation frequency from quasistatic up to 450 kHz. A model based on the equilibrium shape of the meniscus reproduces well the experimental data. Electron microscopy allowed to visualize the meniscus profile around the fiber with a lateral resolution of the order of 10 nm and confirmed its catenary shape. The influence of a lateral confinement of the interface is also investigated. We showed that the lateral extension L of the meniscus influences the effective spring constant following a logarithmic evolution $k \sim 2\pi\gamma/\ln(L/R)$ deduced from the model. This comprehensive study of liquid interface properties over more than four orders of magnitude in meniscus size, shows that advanced FM-AFM and SEM techniques are promising tools for the investigation of mechanical properties of liquids down to nanometer scale.

# INTRODUCTION

Liquid menisci are ubiquitous in many natural and technological processes. For example, they change the physical properties of wet granular media,[1] allow the motion of insects and self-assembly of particles at a liquid interface.[2-4] These surface effects become even more pronounced when scaling down in size.[5] The resulting capillary force may be used to manipulate[6] or fabricate[7-8] 3D objects, whereas it may be detrimental in the development of MEMS and NEMS. It also leads to artifacts in atomic force microscopy (AFM) imaging.[9-10]

Many studies have been devoted to the understanding of the capillary force exerted by a liquid meniscus bridging two solid surfaces,[11] for many relevant geometries and for either volatile[12] or non-volatile liquids. Recently, the development of AFM allowed to probe capillary forces down to nanometer scale, the versatility of AFM modes giving a large variety of solicitations. Static measurements gave information on the liquid nanomeniscus condensed at the tip-substrate contact[13] while non-conventional probes with various types of micro- or nano-fibers at their extremity provided a model geometry to investigate wetting properties at nanoscale.[14-17] The latter case gives access to liquid surface tension by rod-in-free surface type measurements.[18-19] The AFM cantilever can also be used in a passive way to monitor droplet oscillation modes[20] whereas dynamic AFM modes[9, 21] and thermal noise[22-23] also gave access to dissipation processes at stake in the meniscus.

Since the mechanical properties of interfaces are strongly related to their shape, it is also important to develop methods to visualize the liquid interface down to the nanometer scale. Electron microscopy combined with *in situ* micromanipulation techniques now provides a powerful way to tackle this issue. Performing mechanical test in a SEM (Scanning Electron Microscope) adds the advantage of visualizing the experiment, but requires samples that can sustain electron irradiation and above all, vacuum. To study liquids, one has to use nanoaquariums,[24] or if manipulation is needed, liquids with a very low vapor pressure (e.g. ionic liquids IL).[25-26]

In this context, few studies have considered small perturbations around the equilibrium position which leads to the useful description of the liquid bridge as a spring with an effective spring constant mainly depending on contact angle and size.[27-28] Other situations such as droplets or bubbles,[29-30] contact lines[31] or biological membranes[32] have been considered. Interestingly, in most cases, the effective spring constant $k$ depends on the lateral extension $L$ of the liquid interface deformation through $k \sim 1/ln\,(L/R)$ where $R$ is the radius of the probe. However, due to the difficulty to probe a large range in $L/R$, this logarithmic variation has not been assessed experimentally.

Here, we combine two high resolution microscopy techniques (AFM and SEM) in order to study the static and dynamic behavior of a nanomeniscus created by dipping a nanofiber in a liquid interface (Fig. 1). The cylindrical geometry was used to probe both the shape and mechanical properties of the liquid interface down to the nanometer scale. We determined the static and dynamic effective spring constant of the meniscus from experiments performed in the frequency modulation AFM mode (FM-AFM) and showed how this value is altered by a lateral confinement of the meniscus. Interestingly, the use of nanoprobes allowed us to study this effect over many orders of magnitude in lateral size of interfaces. Scanning electron microscopy (SEM) combined with *in situ* nanomanipulation technique was used to directly visualize the shape of the nanomeniscus with high resolution. A theoretical model allows interpreting the results of the experiments on both the shape and the stiffness of nanomenisci, leading to a comprehensive study of the elasticity of liquid interfaces down to nanometer scale.

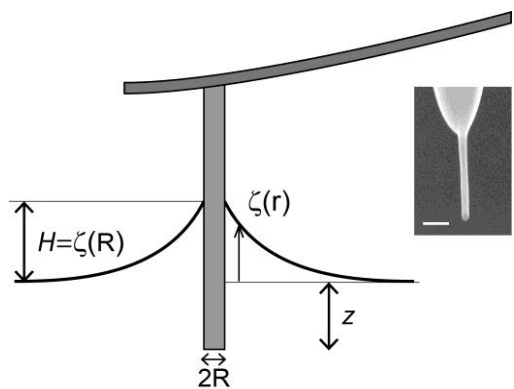

**Figure 1.** Scheme of the experiment. A nanofiber with radius R is dipped in a liquid interface leading to the formation of a meniscus with profile $\zeta(r)$. Inset: SEM image of a 60 nm in diameter silicon nanofiber (CDP55, Team Nanotec); scale bar: 100 nm.

## EXPERIMENTAL SECTION

The experiment consists in dipping in and withdrawing a nanofiber from the free surface of a liquid bath. Two types of tips were used in AFM experiments, both with a nanocylinder *ca.* 50 nm in diameter and *ca.* 500 nm in length at their apex, but with different chemical nature (Silicon probes CDP55 from Team Nanotec and Ag$_2$Ga probes from NaugaNeedles). The spring constant of the cantilevers was chosen of the order of $k_c \sim 2\ N.m^{-1}$, which is much larger than the expected meniscus spring constant while being sufficiently soft to measure forces with good resolution. In our instrument, the nanofiber is inclined by 11° with respect to the vertical. The resulting corrections to the total force coming from the asymmetric deformation of the interface[33] (1,8 %) and from the torque contribution in the AFM measurement[18] (0,4 %) are neglected in the following. In the SEM experiments, we used carbon nanocones[34] terminated by micron scale polyaromatic carbon short fiber which makes the whole object easy to handle with a micromanipulator.[15]

| Name | Cation | Anion | Surface tension (N.m$^{-1}$) at 21°C |
|---|---|---|---|
| IL1-2 | 1-Ethyl-3-MethylImidazolium | Ethylsulfate | 0.0480 |
| IL1-6 | 1-Ethyl-3-MethylImidazolium | Hexylsulfate | 0.0382 |
| IL1-8 | 1-Ethyl-3-MethylImidazolium | Octylsulfate | 0.0355 |
| **IL2-2** | 1-Ethyl-3-MethylImidazolium | Tetrafluoroborate | 0.0560 |
| **IL2-4** | 1-Butyl-3-MethylImidazolium | Tetrafluoroborate | 0.0465 |
| **IL2-6** | 1-Hexyl-3-MethylImidazolium | Tetrafluoroborate | 0.0375 |
| **IL2-10** | 1-Decyl-3-MethylImidazolium | Tetrafluoroborate | 0.030 |

**Table 1.** Name, chemical nature and surface tension of the liquids used.

The choice of liquids was dictated by evaporation issues. Indeed, the level of the liquid interface of a volatile solvent decreases with velocities of the order of the tip velocity.[21] Evaporation may also modify the pinning of the liquid meniscus on the nanofiber, which is essential to analyse the data. Moreover, SEM experiments are performed under high vacuum conditions that are not compatible with the use of standard liquids. We therefore chose to use ionic liquids (ILs), molten salts at ambient temperature[35], with extremely low vapour pressures[36] allowing electron microscopy investigations.[37] We selected two series, differing by the length of the lateral aliphatic chains on either the anion or the cation. For each series, this gives variable surface tension (measured using a Kruss DSA100 goniometer) as reported in Table 1. Note that all experiments were performed in a clean room with controlled humidity. No evolution of the liquid surface tension with time was observed, neither with goniometer nor with AFM. Morever, coulometric titration on IL2-2 left several weeks in ambient condition revealed a limited amount of water in the liquid (2,2 %). The hygroscopy of the ILs is therefore expected to play a minor role.

Under the e-beam of the SEM, some ionic liquids (generally with the heaviest molecular mass) showed a change from liquid to partially solidified behaviour. The measurements of meniscus shapes were carried out on IL2-2 and IL2-4 that remained liquid even after prolonged exposition. Possible solidification under the e-beam has been discussed on ILs similar to the ones used here, mainly for TEM[37-39] for doses ranging from $10^{+19}$ e.m$^{-2}$ to $3.10^{+26}$ e.m$^{-2}$ at acceleration voltages above 120 kV. We performed our SEM observations at a tension of 2 to 5 kV and at doses ranging from $10^{+19}$ e.m$^{-2}$ to $10^{+21}$ e.m$^{-2}$, which means that we stayed on the lower end of the transmitted energy to keep a liquid behaviour.

For AFM measurements, the liquids were put into small cylindrical vessels with various dimensions. A 5 mm large and 2 mm deep hole was drilled in a Teflon sample holder. Since the radius of the hole is larger than the capillary length $\kappa^{-1} = \sqrt{\gamma/\rho g}$ (which is of the order of 2 mm in our experiments), the meniscus can be considered as unbounded (see Section III). Confined liquid interfaces were obtained by milling 100 µm and 10 µm holes in a silicon wafer with a focused ion beam (FIB-SEM Helios NanoLab 600i, FEI). In all cases, the reservoir was filled using a microinjector (Narishige) and the volume of liquid was precisely adjusted in order to get a flat interface pinned on the sharp edges of the reservoir. This is a crucial issue to avoid the formation of a meniscus and the mobility of the contact line at the edge of the reservoir which may strongly modify the mechanical response of the interface.[40]

The AFM experiments were performed on a PicoForce instrument (Bruker) operated in the frequency-modulation mode (FM-AFM) using a phase-lock-loop device (HF2PLL, Zürich Intruments). The nanocylinder at the extremity of the tip was dipped in and withdrawn from the liquid bath, with a typical Z range of 1 µm at a velocity of 2 µm.s$^{-1}$. The capillary force was first measured in static mode from the deflection of the cantilever. Then, the same Z ramp was applied while oscillating the tip at a constant amplitude of 7 nm and monitoring the frequency shift $\Delta f(Z)$.

In order to visualize the shape of the meniscus we also performed experiments in a combined focused ion beam – scanning electron microscope (FIB-SEM, model Helios 600i, FEI). A carbon nanocone[34] was soldered at the extremity of the tungsten tip of a micromanipulator (Omniprobe 200, Oxford Instruments or MiBot from Imina Technologies) using electron- or ion-assisted Pt deposition. Droplets of IL with typical lateral size of 50 µm are introduced in the SEM chamber on a gold wire. The 4 degrees of freedom [x, y, z, and $\theta$ the angle with respect to the (x,y) plane] of the micromanipulator allow to handle precisely the tip and to dip it in the liquid droplet (see Supplementary Information).

## FM-AFM MEASUREMENT OF THE ELASTICITY OF A FREE LIQUID INTERFACE

We first studied the case of an unbounded interface using the large 5 mm container. The nanomeniscus formed around the tip was probed mechanically using the AFM operated in static or dynamic modes as described below:

- For static measurements, the capillary force $F$ is computed from the $z$ cantilever deflection as $F = k_c.z$. An example of curve obtained using a 60 nm in diameter nanocylinder dipped in the free interface of IL2-2 ionic liquid is reported on Fig. 2(a). The interpretation of such curves has already been discussed in several papers.[14-17] The hysteresis between advancing and receding curves is similar to what is obtained macroscopically with a Wilhelmy balance technique. It results from the presence of a large number of defects at the surface of the tip, yet too small to be resolved by SEM. The advancing and receding contact angles are estimated from the mean values of the two flat parts of the curves using the capillary force $F = 2\pi R\gamma cos\theta$.[41] In the case of Fig. 2(a) we find $\theta_{av} = 80° \pm 2°$ and $\theta_{rec} = 68° \pm 2°$. The static stiffness of the meniscus can be determined as $k_{stat} = dF/dz$, provided one only considers sections of the curves where the contact line is pinned, thus insuring a controlled stretching of the interface. Two such sections were identified on Fig. 2(a): when the direction of motion of the tip is reversed (1), due to hysteresis, the contact line remains pinned until the receding contact angle is reached; just before the meniscus snaps off (2), when the nanomeniscus is pinned at the tip apex.[15] The slopes of the *F(z)* measured on these two portions gives the static spring constant with an accuracy of 20 % (including the uncertainty on the determination of the cantilever spring constant which is of order 7 %).

- The same configuration was investigated using the frequency modulation AFM (FM-AFM) mode.[42] In this mode, the cantilever is vibrated at its resonance frequency $f$ using a phase lock loop (PLL) device. A PID modulates the excitation signal $A_{exc}$ in order to maintain constant tip oscillation amplitude. The interaction of the tip with the liquid leads to a frequency shift $\Delta f(Z) = f(Z) - f_0$ where $f_0$ is the resonant frequency in air and to a change in excitation amplitude $A_{exc}(Z)$. The advantage of this mode compared to the standard Amplitude Modulation (AM-AFM) mode used in air, is that it allows to measure independently the conservative (through $\Delta f(Z)$) and the dissipative (through $A_{exc}(Z)$) contributions of the interaction. In this paper, we concentrate on the conservative contribution that shows up as the frequency shift due to the presence of the meniscus. The system can be modeled as a harmonic oscillator with a frequency in air $f_0 = (1/2\pi)\sqrt{k_c/m_c}$ where $m_c$ is the effective mass of the cantilever. A shift in resonance frequency $\Delta f$ may therefore have two origins: (i) a change of spring constant: indeed, the meniscus acts as a spring with stiffness $k_{dyn}$ pulling on the tip which leads to an effective spring constant $k = k_c + k_{dyn}$; (ii) a change in the mass of the system: the oscillation of the nanofiber induces a velocity field in the liquid. The mass of this viscous layer around the fiber then adds to the system. The $\Delta f(Z)$ curve reported on Fig. 2(b) can therefore be interpreted as follows: the sudden positive frequency shift when the meniscus is created is attributed to the meniscus stiffness while the negative slope for positive Z is an added mass effect. Since the added mass effect is negligible at the creation of the meniscus, the associated frequency shift $\Delta f(0)$ reads $\Delta f(0)/f_0 = (1/2)\Delta k/k_c$.[42] The dynamic meniscus spring constant can therefore be determined using $k_{dyn} = 2k_c \Delta f(0)/f_0$ with an uncertainty of order 12 %. The same procedure can be performed using a higher oscillation mode of the cantilever (see Supplementary Information). The frequency of the second mode is given by $f_1 = 6{,}25\, f_0$,[43] whereas its effective spring constant reads $k_{c1} = 39\, k_c$. Note that, since the thickness and therefore the mass of the viscous layer is a function of the liquid viscosity,[44] the negative slope for positive Z is also a function of the liquid viscosity as observed in Fig. 2b.

It is important to note that the pinning of the contact line during the tip oscillation is crucial to guarantee that the meniscus is deformed following the tip motion. This is justified by the fact that the oscillation amplitude (7 nm for all reported data) is much smaller than the distance required to switch from advancing to receding contact angles, observed when the tip motion is reversed [of the order of 100 nm in Fig. 2(a)]. It is also validated by the fact that the frequency shift is independent of the oscillation amplitude up to 21 nm (see Supplementary Information). As discussed above, the use of non-volatile liquid was also dictated by this constraint. These conditions guarantee that the spring constant is measured on a pinned meniscus oscillated at a frequency of the order of 75 kHz for the fundamental mode $f_0$ and 450 kHz for the second mode $f_1$.

The simple cylindrical geometry combined with the use of non-volatile liquids therefore gives a model system with well controlled boundary conditions for FM-AFM measurements. This allows to extract quantitative measurements, whereas in the study of Jai et al[45] evaporation and conical tips were leading to poorly defined menisci.

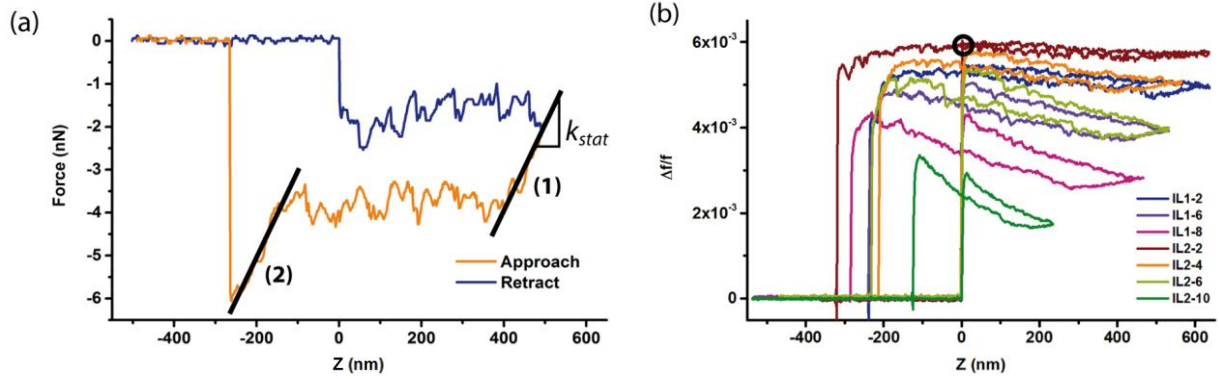

**Figure 2.** (a) Static capillary force measured during immersion (blue line) and retraction (red line) of a nanofiber in a liquid interface. Black straight lines are fits of the two portions of the curves used for the determination of the static spring constant of the interface. (b) Frequency shift as a function of the immersion depth $Z$ for the series of ionic liquids. The black circle shows the $\Delta f(0)$ value used for the determination of the spring constant of IL2-2.

The experimental results are plotted in Fig. 3. We gathered on the same graph all the spring constant values measured in the unbounded meniscus configuration, in static and dynamic modes, at both $f_0$ and $f_1$ frequencies for the latter and with three tips of similar diameter (60 nm) but different chemical nature, two of them in silicon and one in $Ag_2Ga$ alloy.

The spring constant values were plotted as a function of the liquid surface tension. Within the uncertainty of the measure, all data points fall on a same curve, indicating that there is no effect of the frequency of excitation on the response of the liquid interface (up to 450 kHz). This is a strong indication that FM-AFM allows to decouple the effects of interfacial and bulk properties of the liquid which both contribute to the response of the system. Note that this holds for the relatively large viscosities of the liquid used in this study (ranging from 36 mPa.s to 500 mPa.s).

In a first approximation, all data can be fitted by a linear relationship that shows that the spring constant is proportional to the surface tension of the liquid. In the following, we normalize the spring constant values by the surface tension $\gamma$ and defined a reduced spring constant as $k^* = k/\gamma$. We find, for the whole series of liquids investigated, that $k^*_{stat} = k^*_{dyn} = 0.51 \pm 0.06$. The spring constant of an unbounded liquid meniscus is therefore approximately half of the surface tension.

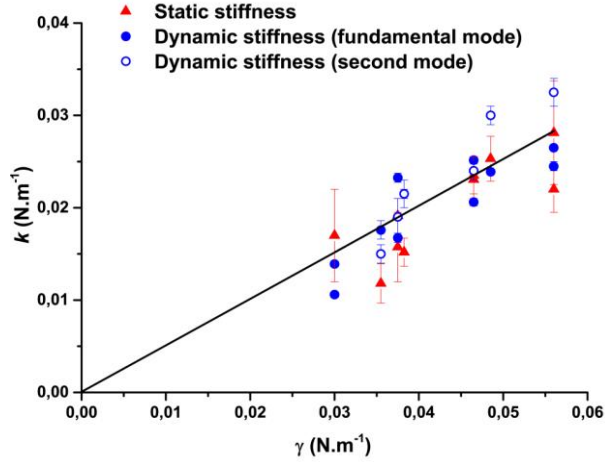

**Figure 3.** Stiffness of the meniscus as a function of liquid surface tension for seven different liquids extracted from deflection signal (static measurement, filled triangles) and from frequency shift (dynamic measurement, open circles for $f_0$ and closed circles for $f_1$).

## THEORETICAL MODEL

### Spring constant of an unbounded liquid interface

In order to interpret these results, we have developed a model to compute the spring constant of a liquid interface. We consider a meniscus pinned on a fiber with a diameter $2R$ dipped in an unbounded liquid bath [Fig. 1 and Fig. 4(a)]. The meniscus spring constant extracted from the capillary force *F* exerted by the meniscus reads

$$k = \frac{dF}{dZ} = \frac{dF}{dH} = -2\pi R \gamma \sin\theta \frac{d\theta}{dH} \quad (1)$$

where *H* is the altitude of the contact line with respect to the flat part of the meniscus i.e. the height of the meniscus (see Fig. 1). The shape of the liquid interface $\zeta(r)$ is solution of the equation

$$\frac{\zeta''}{(1+\zeta'^2)^{3/2}} + \frac{\zeta'}{r(1+\zeta'^2)^{1/2}} = \kappa^2 \zeta \quad (2)$$

which is obtained by a balance, at each point of the interface, of the hydrostatic pressure with the Laplace pressure imposed by the curvature of the meniscus. $\kappa$ is the inverse of the capillary length defined above. It is interesting to note that, in contrast to the 2D meniscus case,[44] Eq.(2) has no analytical solutions. Yet, in the case of a cylindrical fiber, an approximate solution of Eq. (2) was first proposed by Poisson[46] and was later improved by the asymptotic matching technique by Derjaguin[47] and James[48]. The solution $\zeta_{close}(r)$ in the vicinity of the fiber, where gravity [right hand side in Eq. (2)] can be neglected, leads to a catenary profile whereas the solution $\zeta_{far}(r)$ far away from the fiber, obtained for vanishing slopes is a Bessel function (see Supplementary Information). The resulting height of the meniscus on the fiber $H_\infty$ reads:

$$H_\infty = R\cos\theta \left[\ln\left(\frac{4\kappa^{-1}}{R(1+\sin\theta)}\right) - \gamma_E\right] \tag{3}$$

where $\gamma_E \cong 0.57$ is the Euler constant. The ∞ subscript refers to unbounded interfaces with infinite extension. Note that, for fibers with small diameter ($R \ll \kappa^{-1}$), the Euler constant can be neglected as shown in [49]. Combining Eq. (1) and Eq. (3) leads to the expression of the reduced spring constant $k_\infty^*$ for an infinite meniscus:

$$k_\infty^* = \frac{k}{\gamma} = \frac{2\pi}{\ln\left(\frac{\kappa^{-1}}{r_0}\right) - \ln(1+\sin\theta) + \frac{1}{\sin\theta} + \ln 4 - \gamma_E - 1} \tag{4}$$

with the constant $\ln 4 - \gamma_E - 1 \cong -0.191$. The meniscus spring constant is therefore a function of the contact angle of the liquid on the fiber and of the ratio $\kappa^{-1}/R$ of the capillary length to the radius of the probe. A $k^*(\theta)$ curve is plotted in solid black line in Fig. 4(c) for $\kappa^{-1}/R = 66000$, which corresponds to the experimental conditions for an unbounded meniscus. The reduced spring constant $k_\infty^*$ vanishes for a zero contact angle, corresponding to a total wetting case and reaches a nearly constant value for contact angles larger than 30°. Note that the fact that the meniscus stiffness is rather constant validates the description of the interface as a hookean spring. The mean value of the reduced spring constant for a large range of contact angles comprised between 30° and 90° is found to be $k_\infty^* = 0.545$ which is in good agreement with the results reported on Fig. 3.

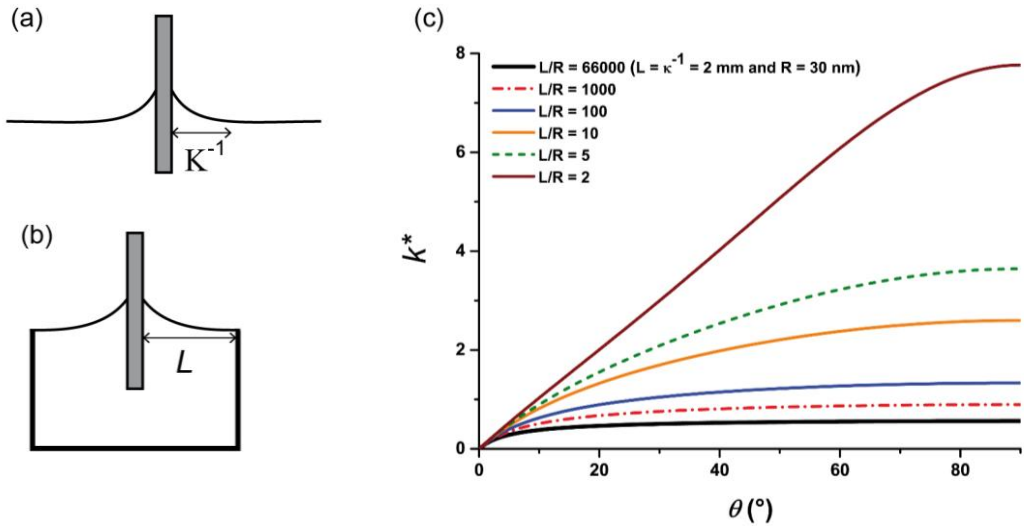

**Figure 4.** (a) Scheme of a fiber dipped in an unbounded liquid interface; (b) Case of a meniscus bounded in a container of size $L$; (c) Plot of the reduced meniscus spring constant as a function of the contact angle $\theta$ for 6 different values of the $L/R$ parameter.

Note that the characteristic time for equilibration of a liquid meniscus can be estimated as $\tau = A/V^*$ where A is the amplitude of deformation and $V^* = \gamma/\eta$ is the capillary velocity of the liquid. In the experimental conditions, we find $1/\tau \sim 50\ MHz$ which is much larger than the FM-AFM excitation frequency used here. Even in dynamic mode, the meniscus can therefore be considered at equilibrium at each instant, consistently with the observation that $k_{stat}^* = k_{dyn}^*$.

*Spring constant of a laterally confined liquid interface*

In this section, we consider the case of a liquid container with a radius $L$ smaller than the capillary length as schematized in Fig. 4(b). The shape of the interface is still given by Eq. (2) but the boundary condition far from the fiber becomes $\zeta(r = L) = 0$, assuming a pinning of the liquid on the sharp edges of the container. Given the nanometric size of our probes, the interface profile can be approximated by a catenary solution (see Supplementary Information) leading to

$$\zeta(r) = R\cos\theta \ln\left[\frac{L+(L^2-R^2\cos^2\theta)^{1/2}}{r+(r^2-R^2\cos^2\theta)^{1/2}}\right] \qquad (5)$$

The height $H_L$ of the meniscus of extension $L$ deduced assuming $\frac{L}{R} \gg \cos\theta$ reads

$$H_L = R\cos\theta \ln\left(\frac{2L}{R(1+\sin\theta)}\right) \qquad (6)$$

Combining with Eq. (1), the expression of the reduced spring constant is obtained:

$$k_L^* = \frac{2\pi}{ln\left(\frac{L}{R}\right)-ln(1+\sin\theta)+\frac{1}{\sin\theta}+ln2-1} \qquad (7)$$

with $ln\,2 - 1 \cong -0.307$. The form of the expression of $k_L^*$ is very close from the $k_\infty^*$ one obtained for unbounded interface [Eq. (4)]. The only difference is in the value of the constant in the denominator which, given the large values of $ln(L/R)$ used, is negligible. Eq. (7) can therefore describe the spring constant of both confined and unbounded liquid interfaces, substituting $L$ by $\kappa^{-1}$ in the latter case. The results of Eq. (7) are reported in Fig. 4 for a large range of $L/R$ values bounded by $L/R = 66000$ which corresponds to the experimental situation reported in Section II. ($L = \kappa^{-1} = 2mm$ and $R = 30\,nm$).

The spring constant of the liquid interface strongly depends on its lateral extension. In a first approximation, this evolution can be described by keeping only the logarithmic term in Eq. (7), leading to an approximate formula:

$$k_L^* \cong \frac{2\pi}{ln\left(\frac{L}{R}\right)} \qquad (8)$$

This logarithmic evolution is similar to the one calculated on bubbles or droplets [29, 50] or contact lines[31] and is characteristic of liquid behaviour.

# COMPARISON WITH EXPERIMENTS

In this section, we report experiments performed by SEM and FM-AFM in order to assess the conclusions of the model described in the theoretical model section.

## SEM observation of nanomenisci

The model presented above is based on a meniscus profile described by a catenary curve [Eq. (5)]. We verified this hypothesis by directly visualizing the shape of the meniscus in a SEM. The experimental procedure involving advanced *in situ* micromanipulation techniques has been described in Section II. Since the nanofibers used in the AFM experiments could not be attached to the SEM micromanipulation tungsten tip, we performed these experiments with homemade carbon nanocones (carbon nanocones terminated by a nanotube[34]).

The motion of the tip was slow enough (about 500 nm/s) to allow SEM monitoring of the process. The motion was stopped a few seconds in order to acquire good resolution images. An example of SEM micrograph of nanomeniscus around a 180 nm diameter carbon fiber is reported in Fig. 5(a). The profile of the meniscus is observed over several microns with a lateral resolution on the order of 10 nm. Since the unperturbed interface profile is not flat due to the use of droplets, we subtracted this latter shape from the measured profile in order to get the influence of the probe on the interface. An example of such a profile extracted with a *MATLAB* program is reported in black in Fig. 5(b). The profile was adjusted by the catenary profile [Eq. (5)] taking the radius of the droplet as $L$. The theoretical profile reproduces perfectly the experimental data with a residual error lower than 1% over 2 µm validating the catenary model [Eq. (5)] which is the basis of the description of the elasticity of a liquid interface presented above.

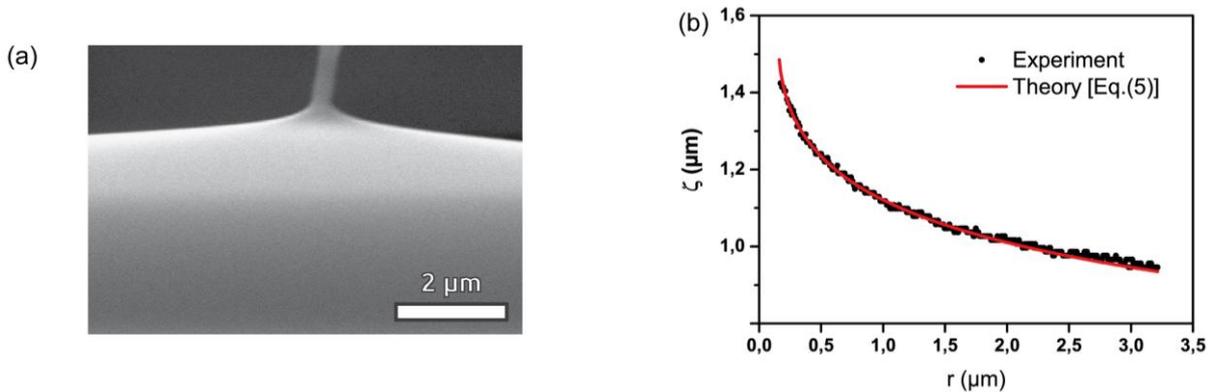

**Figure 5**. (a) SEM image of the meniscus profile created by the carbon nanotube dipped into IL2-4. (b) Meniscus profile and fitted catenary profile [Eq. (5)] as a function of $r$.

## Spring constant of unbounded meniscus

In order to make a quantitative comparison of the experimental values of $k_L^*$ with the model, we first considered the case of unbounded meniscus reported in Section III. In order to take into account the effect of the contact angle, we plotted in Fig. 6, for each experiment, the $k^* = k/\gamma$ value as a function of the contact angle $\theta$ deduced from the measured capillary force. Note that for clarity, we used as $\theta$ the average of advancing and receding contact angles. The results are compared with the theoretical model [Eq. (7)] using $L = \kappa^{-1} = 2mm$ and $R = 30\ nm$.

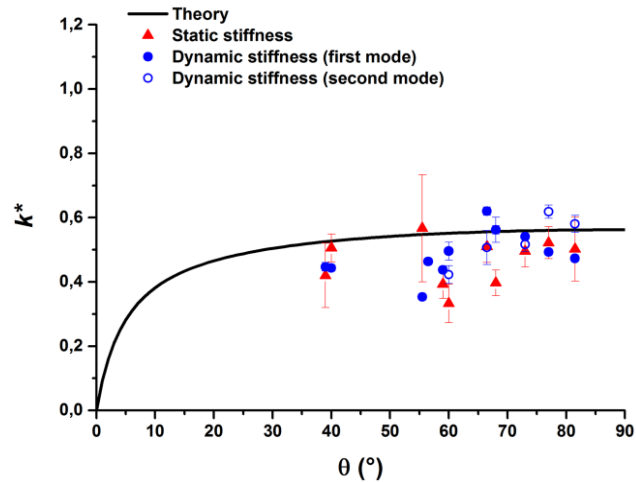

**Figure 6.** Reduced stiffness of the meniscus as a function of liquid contact angle for seven different liquids. The stiffness values are extracted from deflection signal (static measurement) and from frequency shift (dynamic measurement).

The model reproduces fairly well the experimental data: no strong influence of the contact angle is observed in the range of experimental values and the theoretical value (at saturation) $k^* = 0.54$ is consistent with the experimental one. The spring constant of an unbounded meniscus measured with AFM is therefore well described by the formula given by Eq. (7). It shows that the spring constant is proportional to surface tension and can thus be used as a measurement of surface tension of liquids. In particular, the FM-AFM provides a simple and accurate measurement which may be easily automated. Moreover, it does not require detailed control of the pinning of the contact line at the end of the fiber as for rod-in-free surface measurements.[19]

*Spring constant of confined liquid interface meniscus*

*Micron scale containers.* The experimental procedure used on unbounded meniscus was first reproduced in smaller containers with radius $L = 50\ \mu m$ (Fig. 7a) and $L = 5\ \mu m$ (Fig. 7b). For one series of ionic liquids (IL2-2, IL2-4, IL2-6 and IL2-10), we measured the static and dynamic stiffness in the three different containers. The results are reported on Fig. 7c where the spring constant $k$ is plotted as a function of the liquid surface tension $\gamma$ for 2.5 mm (unbounded), 50 µm and 5 µm containers.

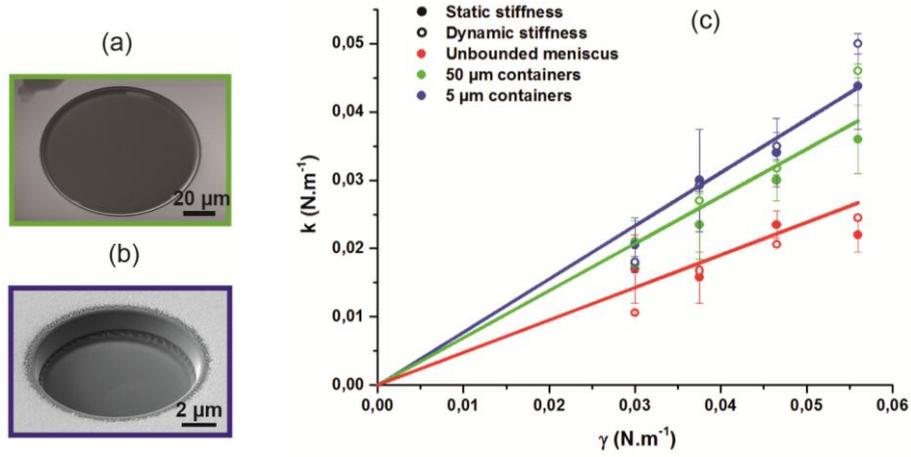

**Figure 7.** (a) SEM image of a container with diameter 100 μm. (b) Same with diameter 10 μm. (c) Stiffness of the meniscus as a function of liquid surface tension for three different containers (unbounded, 50 μm, 5 μm) extracted from deflection signal (static measurement) and from frequency shift (dynamic measurement). The corresponding contact angles, independent of the container size, are comprised between 57° and 82°, leading to a nearly constant reduced spring constant according to Eq.(7).

An increase of the spring constant is clearly evidenced when the size of the liquid interface is reduced. It is a strong indication that in containers the whole interface is oscillated and that, even perturbed very locally with a nanofiber, its deformation extends up to large distances. Since the spring constant is again proportional to the liquid surface tension, we determined reduced spring constants of $k^* = 0.48 \pm 0.1$, $k^* = 0.69 \pm 0.1$ and $k^* = 0.78 \pm 0.1$ for $L/R = 66000$, 1700 and 170, respectively.

*Liquid nanodispensing.* In order to increase the range of $L/R$ values towards smaller values, we also used a method recently developed at CEMES to manipulate ultrasmall liquid quantities on a surface. This nanodispensing (NADIS) technique relies on the use of AFM tips with an aperture at the tip apex (Fig. 8) which connects it to a reservoir droplet deposited on the cantilever.[51-52] By contact of the tip with the surface and retraction, liquid droplets with diameter ranging from 70 nm to several microns are routinely deposited on a surface.[53] In this case, the meniscus extension is limited by the size of the droplet on the substrate. A comprehensive study of the capillary forces experienced by the tip during the deposition process showed that, for nanochannel sizes larger than 200 nm, the curves can be fitted considering that the meniscus is fed by the reservoir during the whole process in order to maintain a zero pressure (or the negligible Laplace pressure of the reservoir).[54] It was also shown that the contact lines remain pinned on the tip and the substrate. These conditions satisfy precisely the assumption of the model, *i.e.* a meniscus pinned on a tip of radius $R$ and on a substrate at a distance $L$, with a negligible pressure. The results can therefore be directly interpreted using the model of Section III.

A static stiffness can be obtained, at each point, as the local slope of the *F(z)* curve *i.e.* $k_{stat} = dF/dz$ (see Supplementary Information). Interestingly, the angle $\theta$ of the liquid interface with respect to the vertical can be deduced, for each point of the curve, using $F = 2\pi R\gamma cos\theta$. We can therefore, with a single experiment, measure a whole curve $k^*(\theta)$ as reported in Fig. 8 in blue dots. In order to increase the range of contact angles above the maximum experimentally available value (30°), we also reported on Fig. 8, in blue dashed line, the modelisation of the capillary force.[54] We observe that the dimensionless spring constant $k^*$, varies from 0 to a value which could be as large as $k^* = 12$ and is limited experimentally to $k^* = 4$. These values are significantly larger than the values obtained for an unbounded meniscus also reported in red in Fig. 8. The $k^*(\theta)$ curve from the NADIS experiment is well described by Eq. (7) using the $L/R = 1.6$ value deduced from the fit of the force curve, as plotted in blue solid line in Fig. 8. This demonstrates that the description of the effective meniscus stiffness provided by Eq. (7) remains valid down to very small $L/R$ values and can therefore be applied to nanomenici.

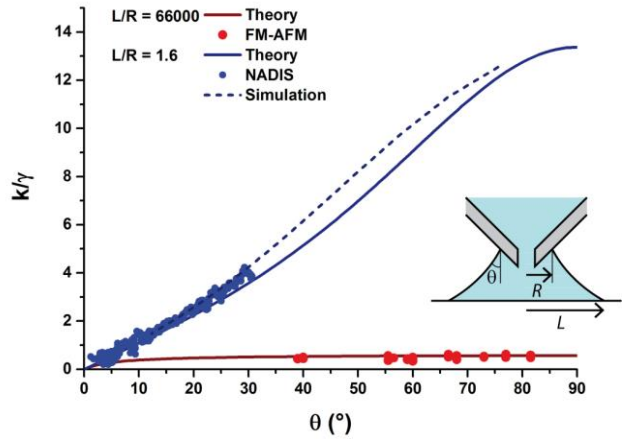

**Figure 8.** Static meniscus spring constant as a function of contact angle: NADIS in blue (experimental (dots) and simulation (dashed line) values deduced from force curve (see Supplementary Information) and theoretical values (solid line) obtained by Eq. (7) with $L/R = 1.6$) and unbounded meniscus in red (Experimental (dots) and theoretical (solid line)). Inset: Scheme of a NADIS tip.

*Effect of lateral extension of the interface.* The four configurations studied experimentally provide a method to study the elasticity of liquid interfaces, with meniscus size ranging from millimetric down to nanometric scale. In terms of the relevant parameter $L/R$ of the model, this covers more than four orders of magnitude, from $L/R = 1.6$ to $L/R = 66000$. We plotted in Fig. 9, the dimensionless spring constant $k^*$ as a function of $L/R$. Note that in the case of NADIS, we considered the value extrapolated using the modelization for a contact angle of 50° which is the mean value of the contact angles obtained in the nanofiber experiments. These results show that the confinement leads to a hardening of the interface by a factor that may be as large as 20 compared to a free unbounded liquid interface.

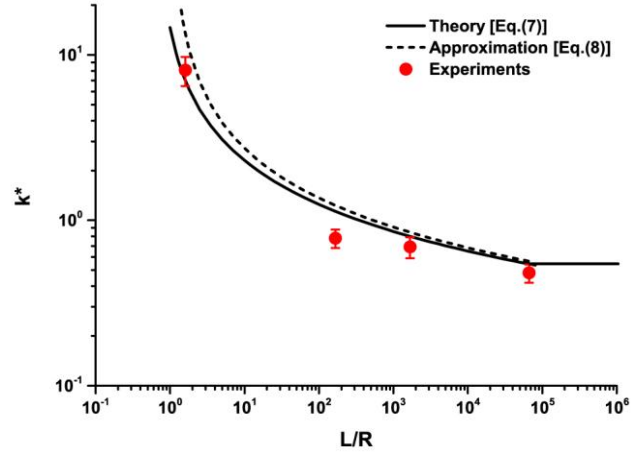

**Figure 9.** Reduced stiffness of the meniscus for an angle $\theta = 50°$ and a tip radius $R = 30\ nm$ as a function of the normalized lateral extension of the liquid interface for NADIS and FM-AFM experiment (unbounded, 50 µm, 5 µm). For $L \geq K^{-1}$, the reduced stiffness is constant and given by Eq. (4).

The results can be directly compared to the theoretical model using Eq. (7) using a contact angle of 50°. This is plotted in Fig. 9 in solid line together with the approximate expression proposed in Eq. (8). This simplified expression which slightly overestimates the spring constant provides a good approximation. The deviation remains smaller than 10 % for $L/R \geq 10$ and, as expected, increases significantly for smaller values. Figure 9 shows that both curves reproduce fairly well the experimental data over about 5 orders of magnitudes in $L/R$ values.

In a fist approximation, the hardening of a liquid interface due to its lateral confinement is inversely proportional to the logarithmic type $(k \sim 1/ln(L/R))$. This is similar to other objects involving liquids interface such as droplets or bubbles[29] or contact line.[31] However, this logarithmic dependence was not verified experimentally. This hardening is qualitatively similar to the one observed with lipid membranes[55] or to the case a solid beam with both ends clamped whose bending spring constant increases when decreasing its length.[56] However, in the latter case the variation of effective spring constant ($k \sim 1/L^3$ for unstressed beam) is more pronounced than for liquids systems.

## CONCLUSIONS

The nanomeniscus created by dipping nanofibers in a liquid interface was investigated using advanced microscopy techniques. We showed that the dynamic FM-AFM mode gives a simple method to measure the dynamic spring constant $k$ of the meniscus. With an unbounded interface, the reduced effective spring constant $k^* = k/\gamma$ was found to be proportional to the liquid surface tension following $k^*_{stat} = k^*_{dyn} = 0.51 \pm 0.06$. We extended this study to the case of interfaces with micron scale lateral dimensions and used non-conventional liquid nanodispensing technique to reach nanometer scale. A strong influence of the meniscus size was evidenced with a 20-fold increase of the spring constant compared to the unbounded interface. A theoretical model based on the equilibrium shape of the interface leads to the expression of the effective spring constant which depends on the liquid contact angle and the ratio of the interface to probe dimensions $L/R$. Neglecting the influence of contact angle, a simple expression $k^* \cong 2\pi/ln(L/R)$ was derived and provides an approximate solution within 10%. The four orders of magnitude range of experimental $L/R$ values available in our experiment, are described by this theoretical model with good agreement. This is, to our knowledge, the first experimental evidence of the logarithmic dependence of the effective spring constant, which was also predicted for other liquid systems. Electron microscopy combined with *in situ* micromanipulation techniques was also used to visualize the shape of the (ionic) liquid interface around the nanoprobe with a 10 nm lateral resolution. The experimental profile, described by a catenary shape within 1%, validates the assumption of the model and demonstrates the potential of EM techniques for manipulation and imaging of liquids down to the nanometer scale.

This study provides a comprehensive description of the shape and mechanical properties of liquid interfaces from macroscopic down to nanometric lengthscales. The new experimental procedures developed to probe nanomeniscus opens many perspectives. The simple relationship between effective spring constant and surface tension provides a new method to measure liquid surface tension by FM-AFM even in a very small volume of liquid (10 pL). Moreover, our work opens the way to the investigation of the dynamical response of complex interfaces involving surface active molecules, or even biological membranes. The use of high frequency AFM modes[57] could even increase the range of available frequency range up to 50 MHz.

In addition to the frequency shift signal which is characteristic of the conservative part of the tip surface interaction, the FM-AFM method can monitor the excitation signal which provides a quantitative measurement of the dissipation processes at stake during the oscillation of the tip. These complimentary data can lead to a full description of the dynamics of nanomeniscus.

The description of interfaces as a spring with well-defined spring constant could be extended to other cases of meniscus in which the capillary pressure becomes predominant. This description can provide a framework to investigate more complicated situations involving a large number of menisci of different sizes as found, for example, in wet granular media or tissues.

## ASSOCIATED CONTENT

*Supporting Information*

Electron microscopy set-up for liquid interface visualization; FM-AFM : measurements at the second resonance mode; FM-AFM : effect of oscillation amplitude; Shape of liquid interface and meniscus height; Capillary forces in liquid nanodispensing.

*Author information*

Corresponding Author : Thierry Ondarçuhu

*E-mail: ondar@cemes.fr

*Acknowledgment*

We thank Elie Raphaël for fruitful discussions. This study has been supported through the French National Research Agency by the NANOFLUIDYN project (grant n° ANR-13-BS10-0009) and, under the "Investissement d'Avenir" program, by the Laboratory of Excellence NEXT (grant n° ANR-10-LABX-0037) and the MIMETIS project (grant n° ANR-10-EQPX-38-01).

# Supplementary information

*Content:*

1. Electron microscopy set-up for liquid interface visualisation

2. FM-AFM: measurements at the second mode

3. FM-AFM: effect of oscillation amplitude

4. Shape of liquid interface and meniscus height

5. Capillary forces in liquid nanodispensing

## 1. Electron microscopy set-up for liquid interface visualisation

The shape of the meniscus was visualized using a combined focused ion beam – scanning electron microscope (FIB-SEM, model Helios 600i, FEI). A carbon nanocone[1] was soldered at the extremity of a tungsten tip of a micromanipulator (Omniprobe 200, Oxford Instruments or MiBot from Imina Technologies) using electron- or ion-assisted Pt deposition. Droplets of IL with typical lateral size of 50 µm are introduced in the SEM chamber on a gold wire. The 4 degrees of freedom [x, y, z, and $\theta$ the angle with respect to the (x,y) plane] of the micromanipulator allow to handle precisely the tip and to dip it in the liquid droplet in a place tangent to the electron beam. This process was repeated several times in order to observe the relaxation of the liquid interface after the meniscus broke up and to verify that no solidification of the liquid was occurring under the electron beam.

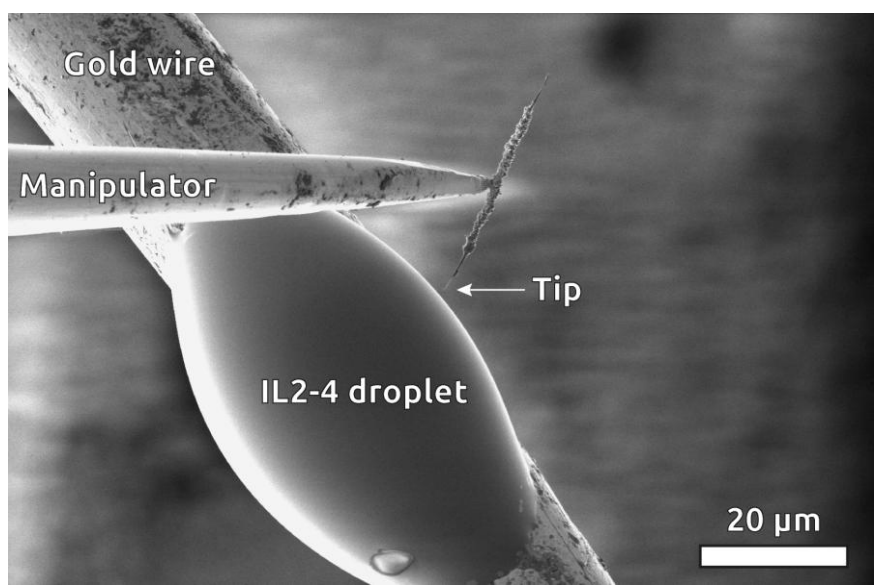

**Figure S1.** SEM image of a carbon tip (up, center of image) attached to the W tip of a micromanipulator (left) approached from a micron size droplet of IL 2-4 deposited on a gold wire.

## 2. FM-AFM: measurements at the second mode

In order to study the influence of oscillation frequency and to reach larger solicitation frequencies, we repeated the FM-AFM experiments, locking the PLL on the second oscillation mode of the cantilever with a frequency $f_1 = 6{,}25 f_0$.[2] The results reported on Fig. S2 show similar trends to the ones observed for the fundamental mode $f_0$ i.e a positive jump resulting from the meniscus formation followed by a decrease of the frequency shift due to added mass effects. The meniscus spring constant is derived from $k_{dyn} = 2k_1 \Delta f(0)/f_1$ where $k_1$ is the effective spring constant of the cantilever corresponding to the second mode which is found to verify $k_1 = 40 k_0$.[2] Note that the curve obtained with the IL2-10 liquid is significantly different for the ones reported in Fig. S2 and is not reported. Excitation and deflection curves also evidence a drastic change in the response of this liquid. Since this liquid is the more viscous one ($\eta = 500 mPa.s$), this may come from the large dissipation associated with the oscillation of this liquid at high frequency and reveals the limits of our procedure.

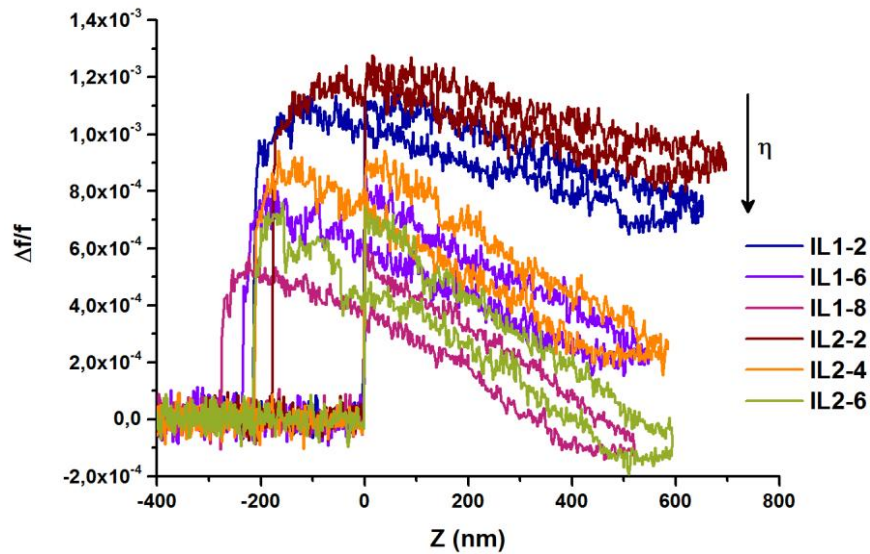

**Figure S2.** Relative frequency shift measured at the second oscillation mode ($f_1 = 450\ kHz$) as a function of the immersion depth *Z* for the series of ionic liquids

## 3. FM-AFM: effect of oscillation amplitude

The experiments reported in the article were performed using small oscillation amplitude $A_{exc} = 7nm$ in order to guarantee that the contact line remains pinned on the tip during the oscillation. In order to verify this, we reproduced the experiments with increasing amplitude (Fig. S3). We observe that the oscillation amplitude has no effect for oscillation amplitudes smaller than $A_{exc} = 21nm$. For larger values, the frequency shift (and therefore the spring constant) starts to decrease. This is due to the fact that the oscillation amplitude becomes larger than the distance required to switch from advancing to receding contact angle. The contact line does not remain anymore pinned on the tip during the whole oscillation. The resulting motion of the contact line leads to a decrease of the effective meniscus stiffness.

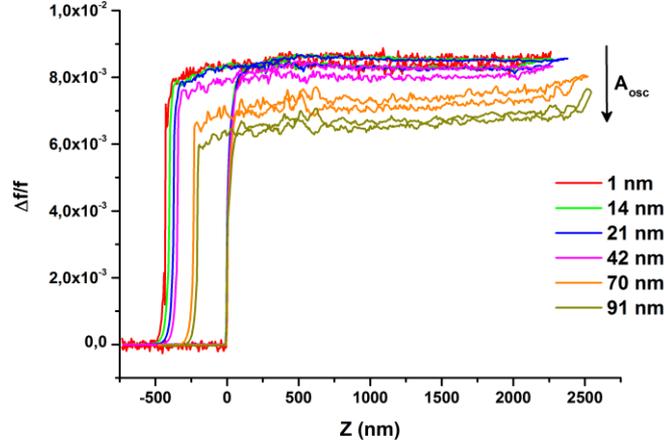

**Figure S3.** Relative frequency shift measured on IL2-2 at the fundamental oscillation mode ($f_0 = 75\ kHz$) as a function of oscillation amplitude $A_{exc}$.

# 4. Shape of liquid interface and meniscus height

The shape of the liquid interface $\zeta(r)$ is solution of the equation

$$\frac{\zeta''}{(1+\zeta'^2)^{3/2}} + \frac{\zeta'}{r(1+\zeta'^2)^{1/2}} = \kappa^2 \zeta \qquad (1)$$

which is obtained by a balance, at each point of the interface, of the hydrostatic pressure with the Laplace pressure imposed by the curvature of the meniscus. $\kappa$ is the inverse of the capillary length defined above.

*Unbounded meniscus*

In the case of an unbounded meniscus, the boundary conditions read $\zeta(r \to \infty) = 0$ and $\zeta'(r = R) = -\cot\theta$. An approximate solution of Eq. (1) based on the asymptotic matching technique in the case of a cylindrical fiber was first proposed by Poisson[3] and derived independently by Derjaguin[4] and James.[5] The solution close to the fiber $\zeta_{close}(r)$, where gravity [right hand side in Eq. (1)] can be neglected, leads to a catenary profile whereas the solution far from the fiber $\zeta_{far}(r)$, obtained for vanishing slopes is a Bessel function.[6] Matching these two approximate solutions leads to:

$$\zeta_{close}(r) = R\cos\theta \left[\ln\left(\frac{4\kappa^{-1}}{(r^2-R^2\cos^2\theta)^{1/2}}\right) - \gamma_E\right] \qquad (2)$$

$$\zeta_{far}(r) = R\cos\theta \cdot K_0\left(\frac{r}{\kappa^{-1}}\right) \qquad (3)$$

where $\gamma_E \approx 0.57$ is the Euler constant. These two solutions are plotted on Fig. S4 together with the numerical solution of Eq. (1) for two different values of the ratio $R/\kappa^{-1}$, namely $R/\kappa^{-1} = 0.1$ and $R/\kappa^{-1} = 1/66000$ which corresponds to fibers with diameter 400 μm and 60 nm, respectively. In the case of a nanoprobe such as the one used in the experiment of the article, the solution $\zeta_{close}(r)$ is a very good approximation up to very large distances compared to the fiber radius (up to $r = \kappa^{-1}/2$), consistent with the fact that gravity can be neglected for such small size probes [see inset Fig. S4(b)].

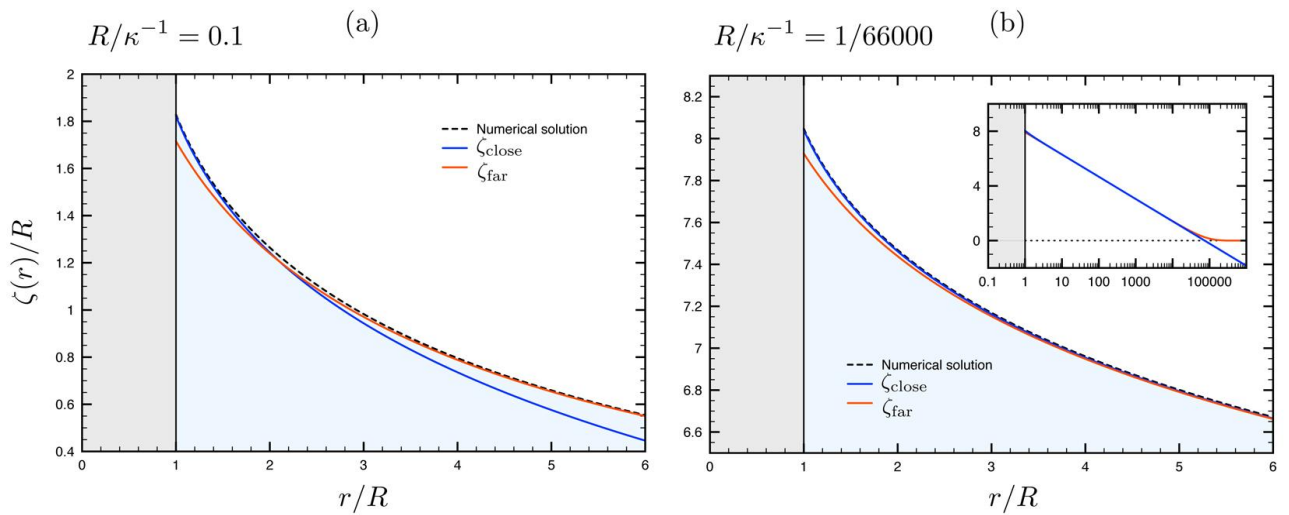

**Figure S4.** (a) Plot of the numerical solution of Eq. (2) (black dashed line) and asymptotic profiles $\zeta_{close}(r/R)$ [Eq. (2), blue solid line] and $\zeta_{far}(r/R)$ [Eq. (3), red solid line] obtained for $R/\kappa^{-1} = 0.1$, which corresponds to a fiber with radius 200 μm. (b) Same for a $R/\kappa^{-1} = 1/66000$, which corresponds to a nanofiber with radius 30 nm; inset: same as Fig. S4(b) with a logarithmic scale for $r/R$.

In all cases, the height of the meniscus on the fiber $H_\infty = \zeta_{close}(R)$ reads:

$$H_\infty = R\cos\theta \left[\ln\left(\frac{4\kappa^{-1}}{R(1+\sin\theta)}\right) - \gamma_E\right] \qquad (4)$$

The $\infty$ subscript refers to unbounded interfaces with infinite extension. Note that, for fibers with small diameter ($R \ll \kappa^{-1}$), the Euler constant can be neglected as shown in ref [7].

*Laterally confined liquid interface*

In the case of a liquid container with a radius $L$ smaller than the capillary length, the shape of the interface is still given by Eq. (1) but the boundary condition far from the fiber becomes $\zeta(r = L) = 0$, assuming a pinning of the liquid on the sharp edges of the container. Given the nanometric size of our probes, as shown in Fig. S4, the interface profile can be approximated by a catenary solution $\zeta_{close}(r)$. The constant is determined using as boundary condition $\zeta(L) = \zeta_{close}(L) = 0$, leading to the interface profile

$$\zeta(r) = R\cos\theta \ln\left[\frac{L+(L^2-R^2\cos^2\theta)^{1/2}}{r+(r^2-R^2\cos^2\theta)^{1/2}}\right] \tag{5}$$

The height $H_L$ of the meniscus of extension $L$ is deduced as $H_L = \zeta(L)$ and, assuming $L/R \gg \cos\theta$ reads

$$H_L = R\cos\theta \ln\left(\frac{2L}{R(1+\sin\theta)}\right) \tag{6}$$

## 5. Capillary forces in liquid nanodispensing

An example of force curve measured during the retraction of a NADIS tip[8] from a surface is reported in Fig. S5. It was obtained with a tip with an aperture of diameter 280 nm. A comprehensive study of the capillary forces experienced by the tip during the deposition process was reported in ref [9]. It showed that, for nanochannel sizes larger than 200 nm, the curves can be fitted considering a constant pressure constraint. In this case, the liquid flow in the channel is sufficient to maintain an equilibrium pressure between the reservoir and the meniscus, at all time. The only adjustable parameter in the model is the $R$ value i.e. the radius of the wetted part of the tip, since $L$ is known experimentally as the radius of the droplet left on the surface after the meniscus breakage. In the case reported Fig. S5, we find $R = 250\ nm$ and $L = 400\ nm$ leading to $L/R = 1.6$. This value was used in the article to calculate the associated spring constant.

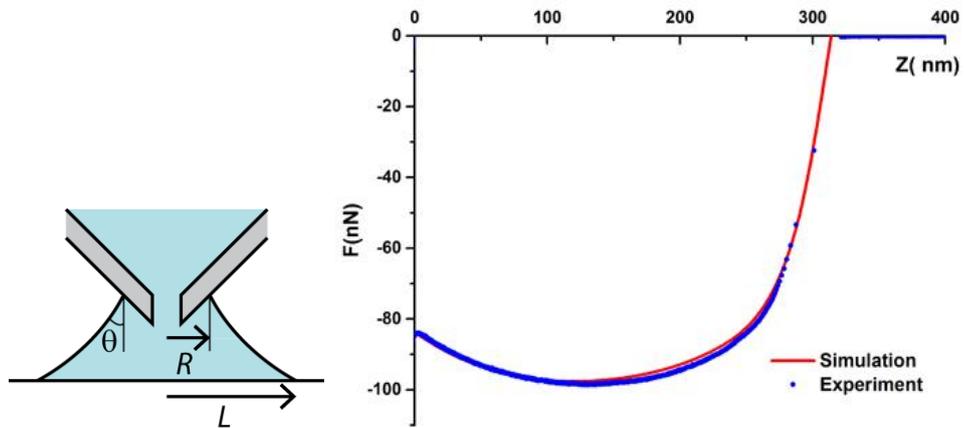

**Figure S5.** (a) Scheme of a NADIS tip; (b) Experimental (blue dots) and calculated (red line) capillary force measured during the retraction of a NADIS tip with a 280 nm diameter aperture from a surface.